\documentclass[prl,twocolumn,showpacs,preprintnumbers,amsmath,amssymb,superscriptaddress, bibnotes]{revtex4-1}

\usepackage{color}
\usepackage{graphicx}
\usepackage{dcolumn}
\usepackage{bm}

\begin{document}

\title{Metamagnetic Behavior and Kondo Breakdown in Heavy-Fermion CeFePO}

\author{S.~Kitagawa}
\email{shunsaku@scphys.kyoto-u.ac.jp}
\author{H.~Ikeda}
\author{Y.~Nakai}
\affiliation{Department of Physics, Graduate School of Science, Kyoto University, Kyoto 606-8502, Japan}
\affiliation{TRIP, JST, Sanban-cho bldg., 5, Sanban-cho, Chiyoda, Tokyo 102-0075, Japan}

\author{T.~Hattori}
\affiliation{Department of Physics, Graduate School of Science, Kyoto University, Kyoto 606-8502, Japan}

\author{K.~Ishida}
\affiliation{Department of Physics, Graduate School of Science, Kyoto University, Kyoto 606-8502, Japan}
\affiliation{TRIP, JST, Sanban-cho bldg., 5, Sanban-cho, Chiyoda, Tokyo 102-0075, Japan}

\author{Y.~Kamihara}
\affiliation{TRIP, JST, Sanban-cho bldg., 5, Sanban-cho, Chiyoda, Tokyo 102-0075, Japan}
\affiliation{Departments of Applied Physics and Physico-Informatics, Keio University, Kanagawa, 223-8522, Japan}

\author{M.~Hirano}
\affiliation{Frontier Research Center, Tokyo Institute of Technology, Yokohama, 226-8503, Japan}

\author{H.~Hosono}
\affiliation{Frontier Research Center, Tokyo Institute of Technology, Yokohama, 226-8503, Japan}
\affiliation{Materials and Structures Laboratory, Tokyo Institute of Technology, Yokohama, 226-8503, Japan}

\date{\today}

\begin{abstract}
We report that nonmagnetic heavy-fermion (HF) iron oxypnictide CeFePO with two-dimensional $XY$-type anisotropy shows a metamagnetic behavior at the metamagnetic field $H_{\rm M} \simeq 4$~T perpendicular to the $c$-axis and that a critical behavior is observed around $H_{\rm M}$. Although the magnetic character is entirely different from that in other Ce-based HF metamagnets, $H_{\rm M}$ in these metamagnets is linearly proportional to the inverse of the effective mass,
or to the temperature where the susceptibility shows a peak. 
This finding suggests that $H_{\rm M}$ is a magnetic field breaking the local Kondo singlet, and the critical behavior around $H_{\rm M}$ is driven by the Kondo breakdown accompanied by the Fermi-surface instability.
\end{abstract}


\abovecaptionskip=-5pt
\belowcaptionskip=-10pt

\maketitle


Metamagnetism is represented by a sudden increase in magnetization with increasing an applied field.
In heavy fermion (HF) systems, CeRu$_{2}$Si$_{2}$ with the tetragonal ThCr$_{2}$Si$_{2}$ structure shows the metamagnetic behavior at about 7.7~T when a magnetic field ($H$) is applied parallel to the $c$-axis.
Although various experiments as well as theoretical studies have been carried out\cite{Y.Hirose_JPCS_2011,Y.Onuki_JPSJ_2004},
the mechanism is still controversial.
In order to understand the metamagnetic behavior in HF systems, it might be desired to investigate new metamagnetic compounds.

The iron oxypnictide CeFePO is a related material of the iron-based superconductor LaFePO\cite{Y.Kamihara_JACS_2006,Y.Kamihara_JPCS_2008}. 
They possess the same two-dimensional layered structure, stacking Ce(La)O and FeP layers alternately. 
Br${\rm \ddot{u}}$ning $et~al.$ reported that CeFePO is a magnetically nonordered HF metal with a Sommerfeld coefficient $\gamma$ = 700~mJ/(mol K$^2$)\cite{E.Buning_PRL_2008}.
At present, it is difficult to synthesize large single crystals of CeFePO for NMR measurements, but $^{31}$P-NMR can probe in-plane and out-of-plane magnetic response separately using $c$-axis aligned polycrystalline samples. 
Here we report novel metamagnetic behavior observed in $H\perp c$, and suggest that metamagnetism of Ce-based HF compounds is driven by Kondo-breakdown (drastic reduction of $c$-$f$ hybridization) as clarified experimentally.

\begin{figure}[tb]
\vspace*{-10pt}
\begin{center}
\includegraphics[width=8cm,clip]{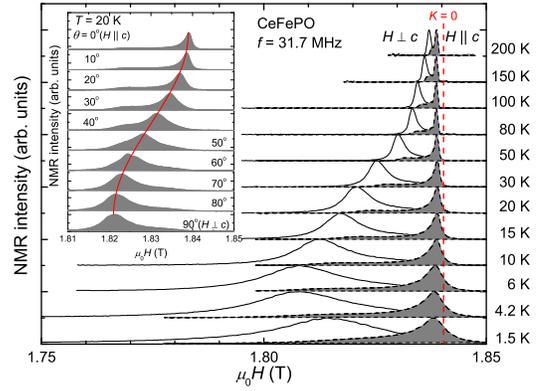}
\end{center}
\caption{(Color online)(Main panel)$T$ dependence of $H$-swept NMR spectra at 31.7~MHz for $H \perp c$ (solid line) and $H \parallel c$ (broken line). $K = 0$ was determined by reference material H$_3$PO$_4$. 
(Inset)Angle dependence of $H$-swept NMR spectra at 31.7 MHz measured at 20 K.
$\theta$ is the angle between magnetic field and $c$-axis. Solid line is corresponding to fitting line.}
\label{Fig.1}
\end{figure}

The polycrystalline CeFePO was synthesized by solid-state reaction\cite{Y.Kamihara_JPCS_2008}.
Basic properties are consistent with the previous report\cite{E.Buning_PRL_2008}.
To measure anisotropic magnetic properties of CeFePO, the samples were uniaxially aligned using a magnetic field\cite{B.L.Young_RSI_2002}. 
The polycrystalline CeFePO was ground into powder, mixed with stycast 1266, and was rotated in the external field of 1.4~T while the stycast cures. 
The $c$-axis of the sample is nicely aligned, which is shown from the angle dependence of $^{31}$P-NMR spectra (see in the inset of Fig.\ref{Fig.1}), and $^{31}$P-NMR measurement was performed on the sample. 

\begin{figure}[tb]
\vspace*{-10pt}
\begin{center}
\includegraphics[width=9cm,clip]{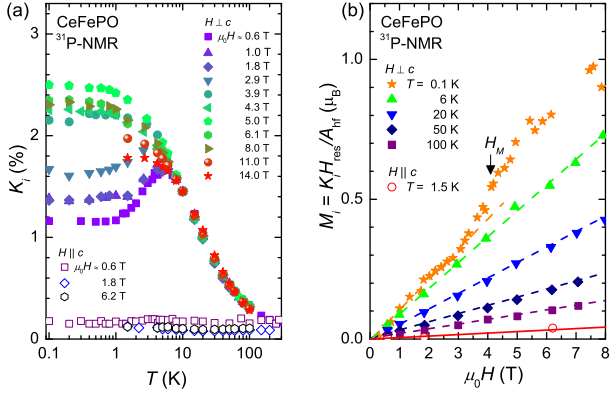}
\end{center}
\caption{(Color online)(a) $T$ dependence of the Knight shift determined at the peaks of $H \perp c$ and $H \parallel c$ spectra obtained at various $H$. 
Strong anisotropy of Knight shift suggests that static spin properties possess $XY$-type spin anisotropy. 
(b) $H$ dependence of magnetization $M_i(H)$ ($i = \perp {\rm and} \parallel$) using the relation of $M_i(H) = K_i(H) H_{\rm res}/ A_{\rm hf}$. Solid and broken lines are guide to eyes. $M_{\perp}(H)$ suddenly increases with increasing $H$ and deviates from linear relation in the field range of 3 - 5~T, which is a definition of a metamagnetic behavior, while such a behavior was not observed in $M_{\parallel}(H)$ up to 6.2~T and down to 1.5~K.}
\label{Fig.2}
\end{figure}

Figure \ref{Fig.1} shows $H$-swept NMR spectra in $H \parallel c$ and $H \perp c$ obtained at 31.4~MHz and various temperatures ($T$). The resonance peak for $H \parallel c$ is almost $T$ independent, but the peak for $H \perp c$ shows the characteristic $T$ dependence originating from $\chi (T)$. 
The Knight shift $K_{\perp(\parallel)}$ was determined from the peak field of the $^{31}$P NMR spectrum obtained in $H$ perpendicular (parallel) to the $c$-axis.
$K = 0$ was determined by reference material H$_3$PO$_4$.
$K_{i}(T, H) (i = \perp {\rm and} \parallel)$, which is the measure of the local susceptibility at the nuclear site, is defined as,
\begin{align}
K_{i}(T, H_{\rm res}) = \left(\frac{H_0-H_{\rm res}}{H_{\rm res}}\right)_{\omega = \omega_0} \propto \frac{M_{i}(T,H_{\rm res})}{H_{\rm res}}
\end{align}
where $H_{\rm res}$ are magnetic fields at resonance peaks, $H_0$ and $\omega_0$ are the resonance field and frequency of bare $^{31}$P nucleus and have the relation of $\omega_0 = \gamma_n H_0$ with gyromagnetic ratio $\gamma_n$, and $M_{i} (T,H_{\rm res})$ is the magnetization under $H_{i,\rm res} (i = \perp {\rm and} \parallel)$ at $T$. 
$K_{\parallel}$ is almost independent of $T$ and $H$, whereas $K_{\perp}$ shows strong $T$ dependence originating from the Curie-Weiss behavior of $\chi (T)$ above 10~K as shown in Fig.\ref{Fig.2}(a). The anisotropic Knight shift suggests that static spin properties possess $XY$-type spin anisotropy. It should be noted that $K_{\perp}$ exhibits $H$ dependence below 4~K and above 2~T, indicative of a non-linear relation between $M_{\perp}$ and $H$.
Using the hyperfine coupling constant $^{31}A_{\rm hf} = 0.2~{\rm T}/\mu_{\rm B}$, which is estimated from the plot between isotropic component of $K$ and  $\chi (T)$ above 10~K (not shown), we can plot $M_{i}(H)$ against $H$ in Fig.\ref{Fig.2}(b). $M_{\perp}(H)$ becomes superlinear against $H$ at 0.1~K, which is the hallmarks of metamagnetism, whereas $M_{\parallel}(H)$ is linear up to 6.2~T, which is again highly anisotropic. 

\begin{figure}[tb]
\vspace*{-10pt}
\begin{center}
\includegraphics[width=8cm,clip]{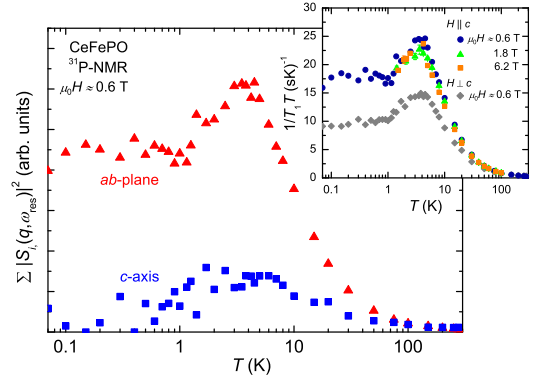}
\end{center}
\caption{(Color online) (Main panel)$T$ dependence of low energy spin fluctuations parallel and perpendicular to the $c$-axis at $\simeq$ 0.6~T evaluated with $1/T_1T$ measured in $H \perp c$ and $H \parallel c$ (See eq. \ref{eq.3}).
The in-plane spin fluctuations are dominant at low $T$, suggesting that the spin dynamics also possess $XY$-type anisotropy. 
(Inset) $T$ dependence of $1/T_1T$ at 10.3~MHz ($\simeq$ 0.6~T) for $H \perp c$ and at 10.3~MHz ($\simeq$ 0.6~T), 31.7~MHz ($\simeq$ 1.8~T), and 107.2~MHz  ($\simeq$ 6.2~T) for $H \parallel c$.
$(1/T_1T)_{H \parallel c}$ is independent of $H$ up to 6.2~T.}
\label{Fig.3}
\end{figure}

Next, we focus on $T$ and $H$ dependence of low-energy spin dynamics probed with the nuclear spin-lattice relaxation rate ($1/T_1$). 
$1/T_1$ of $^{31}$P was measured at each resonance peak by the saturation-recovery method, and was uniquely determined by a single component in whole measured range.
The inset of Fig. \ref{Fig.3} shows $T$ dependence of $1/T_1T$ at low field  $\mu_0H \simeq$ 0.6~T parallel and perpendicular to the $c$-axis. Below 1.5~K, $1/T_1T$ as well as $K$ along both directions becomes constant, indicative of the formation of a Fermi-liquid (FL) state of heavy electrons. In general, $1/T_1$ probes spin fluctuations perpendicular to applied $H$, and thus $1/T_1$ in $H \parallel c$ and $H \perp c$ are described as,
\begin{align}
(1/T_1)_{H \parallel c} &= 2(\mu_0\gamma_n)^2\sum_{q}|H_{\perp}(q,\omega_{\rm res})|^2 \notag \\
&\propto 2A^2\sum_{q}|S_{\perp}(q,\omega \sim 0)|^2, \text{~and} \notag\\
(1/T_1)_{H \perp c} &= (\mu_0\gamma_n)^2\sum_{q}\left[|H_{c}(q,\omega_{\rm res})|^2+|H_{\perp}(q,\omega_{\rm res})|^2\right] \notag\\
&\propto A^2 \sum_{q}\left[|S_{\parallel}(q,\omega \sim 0)|^2 + |S_{\perp}(q, \omega \sim 0)|^2\right].
\label{eq.3}
\end{align}
Here $|X(\omega)|$ denotes the power spectral density of a time-dependent random variable $X(t)$, and $A$ is assumed to be $q$-independent due to the metallic state. 
From these equations, we can decompose spin fluctuations along each direction as shown in the main panel of Fig. \ref{Fig.3}. 
$\sum_{q}|S_{\perp}(q, \omega \sim 0)|^2$ is dominant at low $T$, since $(1/T_1T)_{H \parallel c}$ is almost twice larger than $(1/T_1T)_{H \perp c}$. 
This indicates that the spin dynamics also possess $XY$-type anisotropy. 
The $XY$-type spin fluctuations have the predominance of ferromagnetic (FM) correlations as inferred from the Korringa relation between $(1/T_1T)_{H\parallel c}$ and $K_{\perp}$ in low-$T$ FL state, which is consistent with the previous $^{31}$P-NMR result\cite{E.Buning_PRL_2008} and with the experimental facts that CeFePO is close to FM instability\cite{Y.Luo_PRB_2010,C.Krellner_PRB_2007}. 

\begin{figure}[tb]
\vspace*{-10pt}
\begin{center}
\includegraphics[width=9cm,clip]{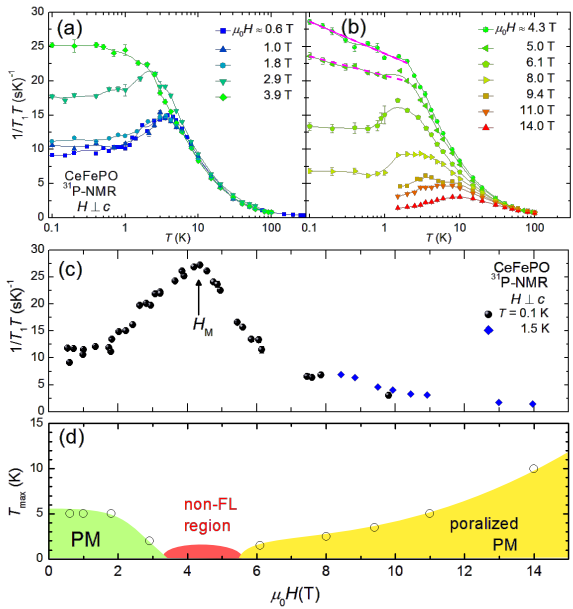}
\end{center}
\caption{(Color online)$T$ dependence of $(1/T_1T)_{H \perp c}$ below 4~T (a), and above 4~T (b).
(c): $H$ dependence of $(1/T_1T)_{H \perp c}$ at low $T$.
$(1/T_1T)_{H \perp c}$ shows a distinct maximum at around $H_{\rm M}$, indicating the enhancement of $N(E_{F})$ related to metamagnetic anomaly. 
(d) $H$ - $T$ phase diagram defined by $T_{\rm max}$ where $(1/T_1T)_{H \perp c}$ shows a maximum. Non FL behavior characterized by continuous increase in $(1/T_1T)_{H \perp c}$ with decreasing $T$ (broken lines are shown in (b)) was observed in a narrow field region intervening between low-field paramagnetic (PM) state and high-field polarized PM state above 6~T.}
\label{Fig.4}
\end{figure}

The evolution of the spin dynamics against $H$ was investigated for both directions. 
Figure \ref{Fig.4} shows $T$ dependence of $(1/T_1T)_{H\perp c}$ below 4~T (approximately the metamagnetic field, $H_{\rm M}$) (a) and above 4~T (b). Although $(1/T_1T)_{H\parallel c}$ does not depend on $H$ up to 6.2~T as shown in the inset of Fig.\ref{Fig.3}, $(1/T_1T)_{H\perp c}$ changes significantly by $H$ as shown in Figs.\ref{Fig.4} (a) and (b). 
$H$ dependence of $(1/T_1T)_{H \perp c}$ at 0.1 and 1.5~K is shown in Fig.\ref{Fig.4} (c). $(1/T_1T)_{H \perp c}$ shows a distinct maximum at $H_{\rm M}$, suggesting that the enhancement of the density of states (DOS) is related to metamagnetic behavior. 
If we assume that $1/T_1T \propto N(E_{\rm F})^2$, $N(E_{\rm F})$ at $H_{\rm M}$ is almost 1.5 times larger than $N(E_{\rm F})$ at 0~T. 
However, it is noteworthy that non FL behavior characterized by continuous increase in $(1/T_1T)_{H \perp c}$ with decreasing $T$ was observed down to 100~mK ($(1/T_1T)_{H \perp c} \sim -\log T$) in 4.3~T $<~\mu_0H~<$ 5~T intervening between low-field paramagnetic (PM) state and high-field polarized PM state above 6~T. 
In the high-field polarized state, $(1/T_1T)_{H \perp c}$ decreases with increasing $H$ and $(1/T_1T)_{H \perp c}$ at 14~T shows almost the same value as (LaCa)FePO [$\simeq$ 1.5 (sK)$^{-1}$] without 4$f$ electrons\cite{Y.Nakai_PRL_2008}.
The $H$ variation of $(1/T_1T)_{H \perp c}$, reflecting the evolution of DOS at the Fermi level with $H$, strongly suggests the evolution of the Fermi surfaces (FSs) by $H$.
It is noted that such a significant $H$ dependence was not reported in the previous specific-heat measurement\cite{E.Buning_PRL_2008}.
This would be because the magnetic field is applied in the various angles against the $c$-axis, and suggests that the metamagnetic behavior would be observed when $H$ is exactly perpendicular to the $c$-axis.

\begin{figure}[tb]
\vspace*{-10pt}
\begin{center}
\includegraphics[width=8cm,clip]{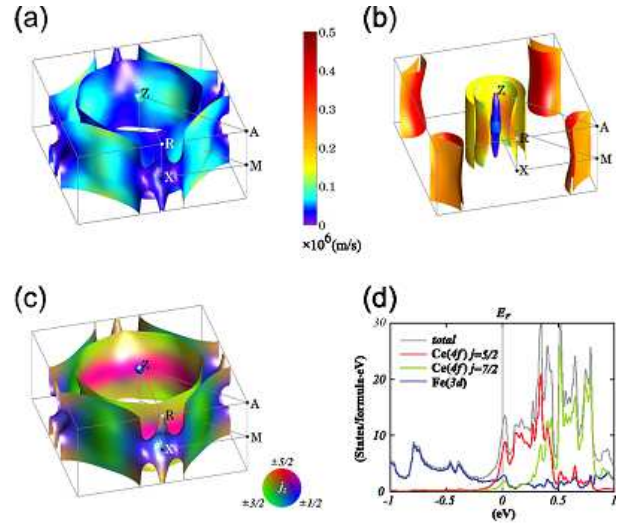}
\end{center}
\caption{(Color online)Fermi surfaces (FSs) calculated with the $ab~initio$ band structure calculation as Ce-4$f$ electrons are itinerant (a), or localized (b), where the Fermi velocity is mapped on the FSs. (c) shows the FS colored by the $j_z$ character in the $j$ = 5/2 multiplet of 4$f$ orbitals, where $j$ is the total angular momentum. (d): the partial density of states. The Fermi level corresponds to 0 eV. }
\label{Fig.5}
\end{figure}

To investigate such evolution of FSs by $H$,
we performed the $ab~initio$ band-structure calculation in the paramagnetic state of CeFePO by using the WIEN2k package\cite{wien2k}.
FSs in the low-field region are composed of itinerant Ce 4$f$ electrons as shown in Fig.\ref{Fig.5} (a). 
The large FS shows the characteristic neck structures around $X$-$R$ at the Brillouin zone boundary, at which boundary electrons have small Fermi velocity or heavy electron mass. The orbital character is dominated by the $j_z$ = $\pm$ 1/2 component in the $j$ = 5/2 multiplet of 4$f$ orbitals, as seen in Figs.\ref{Fig.5} (c) and (d) \cite{cap}. 
These features of Fermi surface imply that the low-field HF state possesses the small $q$ magnetic correlations and their in-plane component is much larger than the out-of-plane component, in good agreement with the experimental results. 
The band calculation also shows that applied field pinches off the neck FSs around $R$ and $X$ in order, that is, ``{\it the field-induced Lifshitz transition}'' appears. 
This is accompanied by a drastic change in DOS at the Fermi level, which can be a driving force for the metamagnetic transition and non-FL behavior around $H_{\rm M}$\cite{T.Senthil_PRB_2008,Y.Yamaji_JPSJ_2006}. 
With applying higher $H$, 4$f$ electrons are localized, and thus the FSs become small. 
The resultant FSs are shown in Fig.\ref{Fig.5} (b), familiar in the iron-based superconductors\cite{S.Lebegue_PRB_2007}. 
Thus the scenario of the field-induced Lifshitz transition can link with the nature in the metamagnetic transition in CeFePO. 

Here, we compare the present results with CeRu$_2$Si$_2$, one of the most well-known metamagnetic compounds. 
Although both compounds show similar $T$ and $H$ dependence of Knight shift and $1/T_1T$ in $H$ parallel to the magnetic easy axis\cite{K.Ishida_PRB_1998}, as well as the similar $H$ - $T$ phase diagram defined by $T_{\rm max}$\cite{Y.Aoki_JMMM_1998} as shown in Fig.\ref{Fig.4} (d), magnetic properties are quite different.
For example, magnetic easy axis is different between CeFePO and CeRu$_2$Si$_2$: CeFePO possesses two dimensional $XY$-type spin anisotropy, whereas CeRu$_2$Si$_2$ possesses Ising-type spin anisotropy\cite{P.Haen_JLTP_1987}. 
As a result, the ground states of the crystal-field level are different and their metamagnetic behavior is observed in different directions.
In addition, dominant magnetic fluctuations in CeFePO differ from those in CeRu$_2$Si$_2$. It is reported that CeRu$_2$Si$_2$ is located close to antiferromagnetic instability accompanied with FM fluctuations\cite{J.Flouquet_LTP_2010,Y.Kitaoka_JPSJ_1985}. 

\begin{figure}[tb]
\vspace*{-10pt}
\begin{center}
\includegraphics[width=7cm,clip]{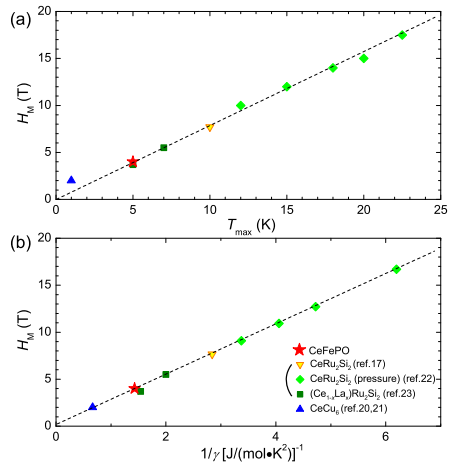}
\end{center}
\caption{(Color online)(a) Values of metamagnetic fields are plotted against $T_{\rm max}$ determined from a maximum in static susceptibility (a) or inverse of Sommerfeld coefficient $\gamma$ at $H$ = 0 (b) in CeFePO, CeCu$_6$ and CeRu$_2$Si$_2$ with pressurized and La-doped system. 
Broken lines are guide to eyes. A linear relation holds between two quantities, although three compounds possess quite different crystal structures and magnetic properties, indicating that $H_{\rm M}$ is linked with the local Kondo singlet energy $T_{\rm max}$.}
\label{Fig.6}
\end{figure}

Figure \ref{Fig.6} shows the relationship between the metamagnetic field $H_{\rm M}$ and the temperature where the bulk susceptibility shows a maximum $T_{\rm max}$ or inverse of Sommerfeld coefficient $\gamma$ at $H$ = 0 for CeCu$_6$, CeFePO, and CeRu$_2$Si$_2$ with doped and pressurized systems\cite{E.Buning_PRL_2008,P.Haen_JLTP_1987,T.Fujita_JMMM_1985,A.Schroder_JMMM_1992,J.M.Mignot_JMMM_1988,R.A.Fisher_LTP_1991}. 
It deserves to note that the linear relation holds between the two quantities, notwithstanding that the three compounds possess totally different crystal structures and magnetic properties. 
Since $T_{\rm max}$ is regarded as a Kondo temperature $T_{\rm K}$ and approximately, the relation of $\gamma T_{\rm K} = const.$ holds in the HF state,
these facts indicate that $H_{\rm M}$ is merely related to the local Kondo singlet energy $T_{\rm max}$ and is not linked with the magnetic fluctuations originating from the intersite coupling between neighboring Ce ions and/or the nesting between the ``large'' FS. 
The experimental facts that $H_{\rm M}$ is linearly proportional to $T_{\rm K}$ in Fig.\ref{Fig.6} strongly suggests that the metamagnetic behavior is linked with the Kondo breakdown\cite{A.Hackl_PRB_2008}. Therefore, in the Ce-based metamagnets, the Kondo breakdown and the Fermi-surface instability accompanied by the drastic change of DOS occur almost simultaneously around $H_{\rm M}$, which can induce novel non-FL behavior.

In summary, we performed $^{31}$P-NMR in the uniaxially-aligned CeFePO and found that CeFePO possesses two dimensional $XY$-type FM fluctuations, and shows metamagnetic behavior when $H$ is applied to $H \perp c$ below 5~K, accompanied with non FL behavior around metamagnetic field $H_{\rm M} \simeq$ 4~T. As far as we know, this is a first example that the metamagnetic behavior occurs in a nonmagnetic Ce-based HF compound with the $XY$-type spin anisotropy. From the band calculation and the comparison with other Ce-based metamagnets, we claim that $H_{\rm M}$ is a magnetic field breaking the local Kondo singlet, which is determined with the intrasite coupling between Ce-4$f$ and conduction electrons, and that the FSs change drastically due to the Kondo breakdown.

This work was partially supported by Kyoto Univ. LTM center, the "Heavy Electrons" Grant-in-Aid for Scientific Research on Innovative Areas  (No. 20102006) from The Ministry of Education, Culture, Sports, Science, and Technology (MEXT) of Japan, a Grant-in-Aid for the Global COE Program ``The Next Generation of Physics, Spun from Universality and Emergence'' from MEXT of Japan, a grant-in-aid for Scientific Research from Japan Society for Promotion of Science (JSPS), KAKENHI (S and A) (No. 20224008 and No. 23244075) and FIRST program from JSPS. One of the authors (SK) is financially supported by a JSPS Research Fellowship.



%

\end{document}